\documentclass{aa}  

\usepackage{graphicx}
\usepackage{color}

\newcommand{\Msun} {$\mbox{M}_{\sun}$}

\newcommand{\sthreeD}{$\sigma_{\mathrm{3D}}$}

\usepackage{gensymb}
\usepackage{txfonts}

\usepackage[usenames,dvipsnames,svgnames,table]{xcolor}
\usepackage[breaklinks, colorlinks, citecolor=CornflowerBlue]{hyperref}
\usepackage{url}
\begin{document} 

\newcommand{\ct}[1]{\textcolor{brown}{#1}}

   \title{Universal gravity-driven isothermal turbulence cascade in disk galaxies} 
\titlerunning{Isothermal Turbulent Cascade}

   \author{J\'er\'emy Fensch\inst{1},
        Fr\'ed\'eric Bournaud\inst{2},
        No\'e Brucy\inst{2,3},
        Yohan Dubois\inst{4},
        Patrick Hennebelle\inst{2} and Joakim Rosdahl\inst{1}
          }
   \authorrunning{Fensch J. et al.}
    \offprints{Jérémy Fensch (jeremy.fensch@ens-lyon.fr)}

   \institute{
  $^1$ Univ. Lyon, ENS de Lyon,  Univ. Lyon1, CNRS, Centre de Recherche Astrophysique de Lyon UMR5574, F-69007 Lyon, France \\
  $^2$ Université Paris-Saclay, Université Paris Cité, CEA, CNRS, AIM, 91191, Gif-sur-Yvette, France \\
  $^3$  Universität Heidelberg, Zentrum für Astronomie, Institut für Theoretische Astrophysik, Albert-Ueberle-Str 2, D-69120 Heidelberg, Germany \\
  $^4$ Institut d'Astrophysique de Paris, CNRS, Sorbonne Universit\'{e}, UMR7095, 98bis bd Arago, 75014 Paris, France \\ }

   \date{Submitted Nov. the 17th 2023, Accepted Jan the 30th 2023}

 
  \abstract{
While interstellar gas is known to be supersonically turbulent, the injection processes of this turbulence are still unclear. Many studies suggest a dominant role of gravitational instabilities. However, their effect on galaxy morphology and large-scale dynamics vary across cosmic times, in particular due to the evolution of the gas fraction of galaxies. In this paper, we propose numerical simulations to follow the isothermal turbulent cascade of purely gravitationaly-driven turbulence from its injection scale down to 0.095~pc for a gas-poor spiral disk and a gas-rich clumpy disk. To this purpose, and to lift the memory-footprint technical lock of sufficiently resolving the interstellar medium of a galaxy, we developed an encapsulated zoom method that allows us to probe self-consistently the self-generated turbulence cascade over three orders of magnitude on spatial scales. We follow this cascade for 10~Myrs. We find that the turbulent cascade follows the same scaling laws in both setups. Namely, in both cases the turbulence is close to equipartition between its compressive and solenoidal modes, the velocity power spectrum follows the Burgers' scaling and the density power spectrum is rather shallow, with a power-law slope of -0.7. Last, gravitationally-bound substructures follow a mass distribution with a -1.8 slope, similar to that of CO clumps. These simulations thus suggest a universality of gravity-driven isothermal turbulent cascade in disk galaxies across cosmic time. }
               
   \maketitle
%
\section{Introduction}
\label{sec:intro}

    Gas in galaxies is supersonically turbulent \citep[see review by][]{Elmegreen2004, Hennebelle2012}. The turbulent energy decays on a timescale much shorter than the time for which molecular clouds are turbulent. Thus, there must be a mechanism to continuously provide turbulent energy to molecular clouds. 
    
    Processes suggested so far have been shear from galactic rotation \citep{Fleck1981}, mass transport \citep{Krumholz2016, Krumholz2018} stellar feedback \citep[see e.g.][]{MacLow2004, Ostriker2011, Faucher-Giguere2013, Padoan2016, Hayward2017}, gravitational instabilities \citep[see e.g.][]{Bournaud2010} and gas accretion \citep{Klessen2010, Goldbaum2015, Goldbaum2016, Ginzburg2022, Forbes2022}. However, distinguishing their respective roles is not an easy task. 
    
    For instance, \citet{Krumholz2016} developed an analytical prediction for the dependence of the gas velocity dispersion on star formation rate in galaxies with gravity- and feedback-dominated turbulence injection. They found that observational data is best fitted by gravity-dominated injection for high-redshift galaxies, whereas the low-redshift case is less clear, predictions of the two models being very close to each other \citep[see also ][]{Varidel2020, Girard2021, Yu2021}. Using numerical simulations with self-gravity and stellar feedback, \citet{Bournaud2010} found that the turbulent cascade characteristic length, for a Large Magellanic Cloud-type galaxy model, is set by the Jeans length, and that gravity dominates turbulence injection. Using simulations of 1 kpc$^{3}$ sized regions of galaxies with turbulent forcing, \citet{Brucy2020} showed that turbulence injection via feedback may be sufficient to regulate star formation for gas surface densities typical of $z \simeq 0$, but stronger driving becomes necessary at larger gas surface densities. These theoretical studies thus suggested that gravitational instabilities may be the most important injection mechanism, while others suggested stellar feedback remains dominant, at least for $z\simeq0$ \citep[see e.g.][]{Orr2020}. \\
    
    On the observational side, it has been observed that galaxy-wide velocity dispersions are well correlated to their star formation rates, thus implying a connection between turbulence and star formation feedback \citep[e.g.][]{Green2010,Lehnert2013}. However, some giant molecular clouds (GMC) are seen to be turbulent even without star formation \citep[see e.g.][]{Poidevin2013}. For instance, observing 272 GMCs in CO(1-0) in the Large Magellanic Cloud, \citet{Kawamura2009} did not find a different linewidth between clouds with no massive star formation, those with H{\sc II} regions and those with H{\sc II} regions and young clusters. Furthermore observations of gas-rich local galaxies \citep[][]{Fisher2017} and $z\simeq2$ galaxies \citep[see e.g.][]{Ubler2019}, found that both categories of galaxies are marginally \citet{Toomre1964}-stable, suggesting a predominant role of gravity in their elevated velocity dispersion. It is thus necessary to study these two processes independently to better understand their respective roles.

    However, the turbulent injection process is likely to have evolved among cosmic times. In particular, the high gas fraction observed in $z\simeq2$ galaxies should change drastically the morphology and internal dynamics of galaxies compared to galaxies in the local Universe, which tend to have low gas fractions \citep[see ][and references therein]{Fensch2021}. Gravity-driven turbulence and its cascade down to star formation region scales may thus be a function of gas fraction.
    
    In this study we propose to tackle this question by following the turbulent cascade generated purely by gravitational instabilities and galactic dynamics for gas-poor disks and gas-rich disks, under the simplifying assumptions of an isothermal equation of state and no star formation nor stellar feedback. Thus, except for the isothermal equation of state, our numerical methods do not include sub-grid models. Our purpose is to characterize the gravity-generated turbulence cascade and to investigate whether this cascade changes with the gas fraction of the galaxy. The study of the interplay between gravity and star formation feedback will be presented in a forthcoming publication.

    The numerical methods and simulation sample are presented in Section~\ref{sec:simulation}. Results are described in Section~\ref{sec:results}. The discussion and conclusions are presented in Section~\ref{sec:discussion} and \ref{sec:conclusions}, respectively. The Appendix presents the study of the robustness of the different results with respect to the chosen numerical methods. In what follows, the number density $n$ is defined as the number density of hydrogen nuclei, which is assumed to have a mass fraction of 0.76 and $\rho$ is defined as the total mass density, including both hydrogen and helium. Pressure is derived from temperature using the ideal gas law and hydrogen and helium densities, with a mean molecular weight given by the ionisation state of the gas.

\section{Simulation code and disk models}
\label{sec:simulation}
    
    In what follows, we describe our numerical methods and the idealized disk models we use.
    
    \subsection{Numerical methods}
    \label{subs:method}
    
    We use the {\sc ramses} simulation code \citep{Teyssier2002}. {\sc ramses} uses adaptive mesh refinement (AMR), and solves the hydrodynamics equations using a second-order Godunov method (MUSCL scheme) and solves the Poisson equation using a conjugate gradient method. We use an isothermal equation of state for gas denser than $10^{-3}$~cm$^{-3}$ , with temperatures of 236~K or $10^4$~K, depending on the galaxy models described in the following subsection. Gas less dense than $n = 10^{-3}$~cm$^{-3}$ is set to the virial temperature $T = 4 \times 10^6  ( n / 10^{-3}~\mathrm{cm}^{-3} ) ^{2/3} $ ~K \citep{Bournaud2010} to maintain the halo gas in gravothermal equilibrium. The reason for using a simple isothermal equation of state is to limit ourselves to purely gravitational instabilities, and not the thermal ones. The effect of a full cooling model is discussed in Section~\ref{sec:disc:physics}.

    Cells are refined whenever they contain more than 50 initial condition particles, their mass (gas plus particles) exceeds $5\times10^{4}$~\Msun, or when the Jeans length is less than four cell width. The box size is 100~kpc$^{3}$. The coarsest cells have a width of 781~pc (level 7) and refinement is allowed to a width of 6~pc (level 14). For gas at the maximum resolution for which the local Jeans length is less than four cell widths, we increase the temperature of the cell following a polytrope equation, with T $\propto~\rho$ providing a pressure floor preventing artificial fragmentation, which we will refer to as the \citet{Truelove1997} criterion \citep[see][for details]{Teyssier2010}. 
    
    After 100~Myr of evolution, we chose by eye an over-density region in each simulation, at approximately the same galacto-centric radius of 3~kpc, and we store its orbit. The region is a sphere of 1~kpc diameter, that will be called the 'zoomed region' hereafter. We re-run the simulation and this time allow more refinement in the zoomed-region which follows the stored orbit. We note that the extra refinement in the zoom region is only seen by the gas and not by the particles, which only see refinement up to the previous 6 pc resolution to prevent over-sampling their gravity. The refinement is allowed to AMR level 20, that is down to 0.095~pc. The refinement strategy follows the same criteria as above, except that the refinement is also triggered whenever the local Jeans length is less than 30~cell widths, which is the convergence criterion to resolve the solenoidal motions of the turbulence presented by \citet{Federrath2011}. Furthermore, a maximum cell width of 3~pc is imposed for all cells less than 500~pc from the center of the zoomed region.
    
    To avoid numerical artefacts due to the refinement timescale, e.g. the creation of spurious shocks if the gas does not have time to settle at the new resolution limit, we activate a new level of refinement every one Myr, which is much larger than the timescale for convergence found by \citet{Seifried2017}. 
    
\begin{figure}
        \centering
        \includegraphics[width=9cm]{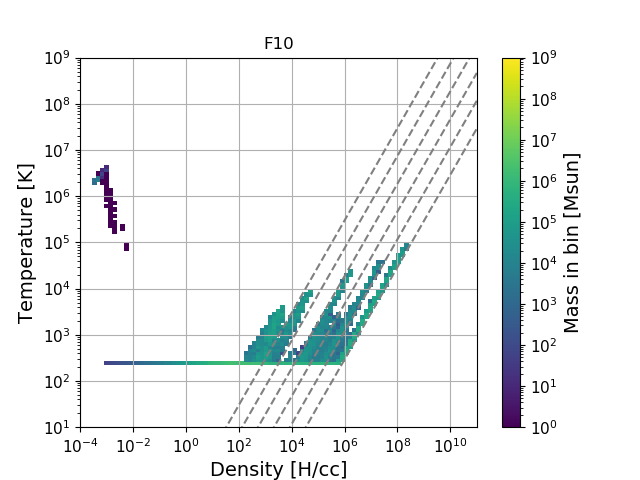}
        \includegraphics[width=9cm]{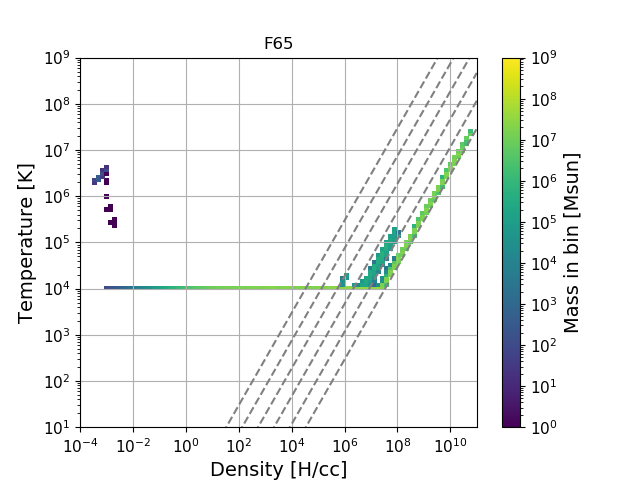}
        \caption{ Density-temperature diagrams for the zoom regions, 10~Myr of evolution after the activation of level 20. The dashed lines show the minimum pressure floor to satisfy the \citet{Truelove1997} criterion, for levels 15 to 20, from left to right. This pressure floor is smoothly reduced towards the center of the zoom (see text). } 
        \label{fig:rho_T}
    \end{figure}

        Furthermore, in order to prevent collapse of the gas entering the zoomed region, we use a 'target strategy': level 15 (corresponding to a cell width of 3 pc) is triggered only within 600 pc from the centre of the zoomed region, level 16 within 550 pc, and so on, up to level 20 within 350 pc. To prevent any jump in the pressure floor (see above) it evolves linearly such that the Jeans length is always larger than four cell width at the highest resolution. At the maximum level, the Jeans polytrope is activated for densities above $10^6 \,\rm H\,cm^{-3}$ and $3.5\times10^7 \,\rm H\,cm^{-3}$ for isothermal temperatures of 236~K and $10^{4}$~K, respectively. The density-temperature diagrams, corresponding to the zoom regions, are shown in Fig.~\ref{fig:rho_T}. In the zoom regions presented in this paper, we reach up to 26 millions leaf cells (i.e. cells that are not split into sub-cells).

     We stress that we do not extract the region in the zoom for a new simulation, as is often done in the literature (see e.g. \citealt{vanLoo2013, Bonnell2013, Butler2015, Dobbs2022}, but see \citealt{Smith2020}, with an analytical background gravitational potential). Instead, in our setup the full galaxy is simulated self-consistently along with the zoom region, in order to retain shear and large-scale gravity.

    \subsection{Simulation set}
    \label{subs:set}

   We use the G2 galaxy model from \citet{Perret2014}, which is based on the MASSIV sample of $z \sim 1.5$ galaxies \citep{Contini2012}. The parameters of the simulations are described in Table~\ref{table::galaxies}. We study two disk galaxies, F10 and F65 with the same total mass distribution, thus the same rotation curve. The gaseous and stellar disks are initialised with an initial Toomre $Q$ parameter of 1 and 1.7, respectively. The two galaxies differ in two ways:

\begin{itemize}
    \item (i) the F10 (resp. F65) galaxy has a gas mass fraction of 10\% (65\%). Gas fraction is defined as gas mass over the sum of the the stellar and gas masses.
    \item(ii) the value of the constant temperature is chosen as to get a similar $Q_\mathrm{gas}$ in both cases. We chose to do this to isolate the effect of changing the gas fraction. Given that the shear curve is set to be the same, that the sound speed $c_{\rm s}$ goes with the square root of the temperature, and that the gas surface density is 6.5 times higher in the F65 case, we set T$_{F10}$ = T$_{F65}$/$6.5^{2}$. We chose T$_{F65} = 10^4$~K \citep[see e.g.][]{Behrendt2016}, and thus T$_{F10} = 236$~K.

\end{itemize}

    \begin{table}
    \centering
    \caption{Characteristics of the simulated galaxies. The relative gas and stellar masses are chosen to have a baryonic gas mass fraction of 10\% and 65\% for F10 and F65, respectively. The relatively low mass of the dark matter halo comes from the radial truncation of the halo.}
    \label{table::galaxies}
    \begin{tabular}{l||cc}
    \hline 
    Galaxy                                              & F10                 & F65                      \\ \hline
    Total baryonic mass [$\times 10^{10} $M$_{\odot}$]  & \multicolumn{2}{c}{4.46}                                                                  \\ \hline
    {\bf Gas Disc (exponential profile)}                               &                      &                              \\
    mass [$\times 10^{10} $M$_{\odot}$]                        & 0.46                   & 2.9                            \\
    characteristic radius [kpc]                         & \multicolumn{2}{c}{2.6}                                                                   \\
    truncation radius [kpc]                          & \multicolumn{2}{c}{8.0}                                                                     \\
    characteristic height [kpc]                      & \multicolumn{2}{c}{0.15}                                                                  \\
    truncation height [kpc]                           & \multicolumn{2}{c}{0.45}                                                                   \\  \hline

    {\bf Stellar disc (exponential profile) }                           &                      &                                    \\
    mass [$\times 10^{10} $M$_{\odot}$]                           & 4.02                   & 1.42                        \\
    characteristic radius [kpc]                         & \multicolumn{2}{c}{1.6}                                                                   \\
    truncation radius [kpc]                          & \multicolumn{2}{c}{8.0}                                                                     \\
    characteristic height [kpc]                      & \multicolumn{2}{c}{0.30}                                                                  \\
    truncation height [kpc]                           & \multicolumn{2}{c}{1.5}                                                                   \\  \hline

    {\bf Bulge (Hernquist profile) }                               &                      &                                 \\
    mass [$\times 10^{10} $M$_{\odot}$]            & \multicolumn{2}{c}{0.14}                 \\
    characteristic height [kpc]                      & \multicolumn{2}{c}{0.32}                                                                  \\
    truncation height [kpc]                           & \multicolumn{2}{c}{1.6}                                                                   \\  \hline
    {\bf Dark Matter halo (Burkert profile) }                      &                      &                                     \\
    mass [$\times 10^{10} $M$_{\odot}$]            & \multicolumn{2}{c}{14.48}                 \\
    characteristic radius [kpc]                      & \multicolumn{2}{c}{15.0}                                                                  \\
    truncation radius [kpc]                           & \multicolumn{2}{c}{35.0}                                                                   \\  \hline

    \end{tabular}
    \end{table}

    \begin{figure*}[h!]
        \centering
        \includegraphics[width=17 cm]{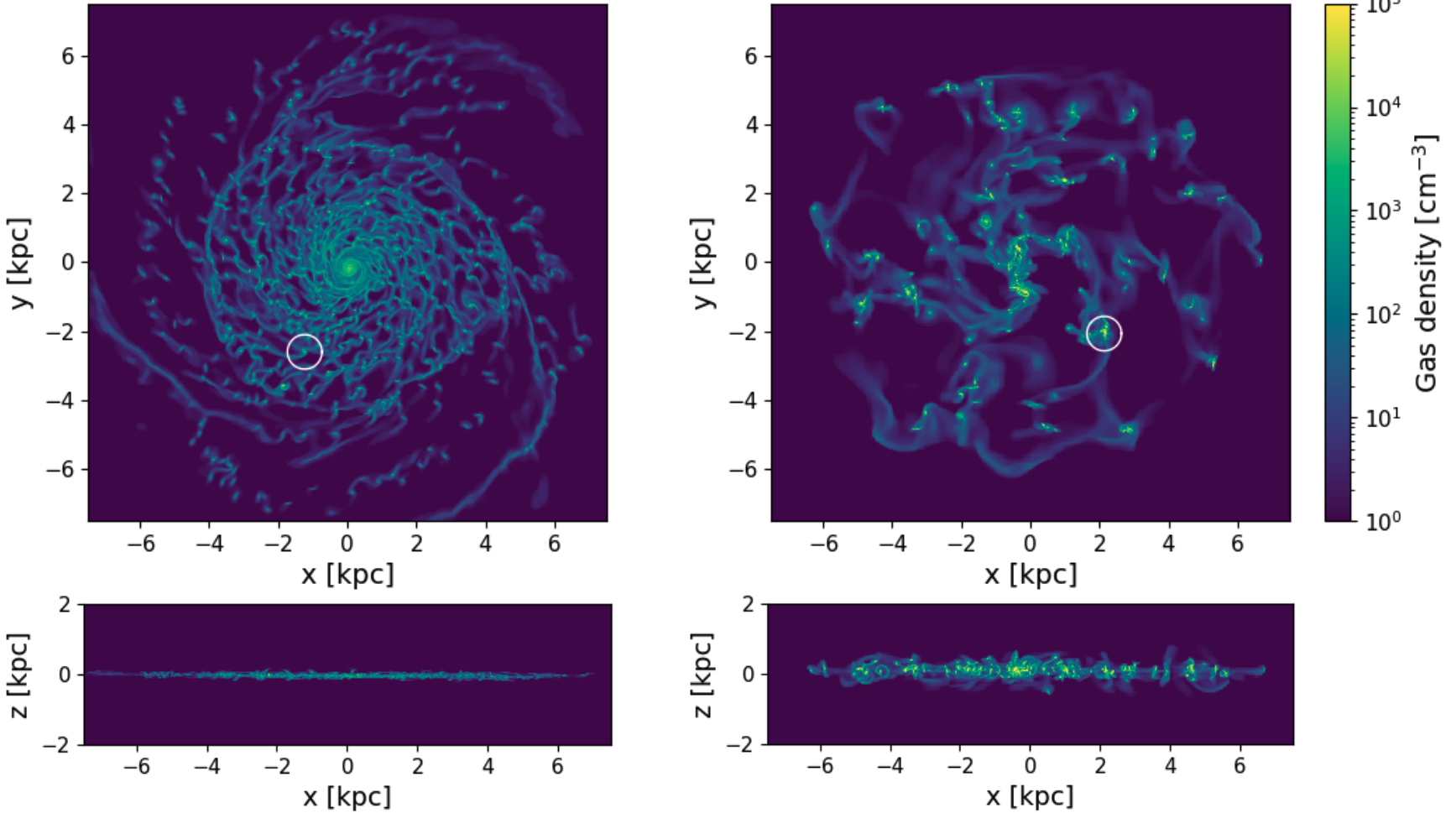}
        \caption{Gas density of the F10 (left) and F65 (right) runs. The white circles show the location of the zoom region.}
        \label{fig:star_surf_dens}
    \end{figure*}

    \subsection{Disk model characteristics}
         
    The gas and stellar density maps of the simulations before starting the zoom procedure are shown in Fig.~\ref{fig:star_surf_dens}. The F10 model develops a spiral morphology, both in its gaseous and stellar components, and has a gas disk height of around 100~pc. 
    The F65 model develops a clumpy morphology both in its gaseous and stellar components, and has a gas disk height of around 1~kpc. This is similar to what is obtained for other isothermal gas-rich disk simulations \citep[see e.g.][]{Behrendt2016}. \\

        \begin{figure}
        \centering
        \includegraphics[width=9cm]{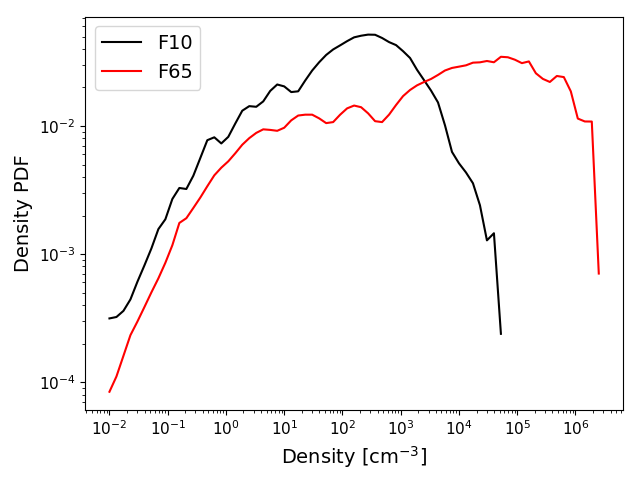}
        \caption{Normalized gas density PDFs for the F10 and F65 runs at the time shown in Fig~\ref{fig:star_surf_dens}, i.e. before the refinement in the zoom-in region is triggered. Local maxima in the density PDF correspond to density thresholds for refinement.}
        \label{fig:no_zoom_PDF}
    \end{figure}

      In Fig.~\ref{fig:no_zoom_PDF} we show the density probability density functions (PDF) in the simulations on the snapshots illustrated in Fig.~\ref{fig:star_surf_dens}. From turbulence theory, the density PDF of isothermal fluids with supersonic turbulence is expected to be well represented by a log-normal form \citep[see e.g.][]{Kritsuk2007, Federrath2008, Federrath2010}. We resolve densities up to a few $10^4$ and $10^6$ H/cc for F10 and F65, respectively.

        \begin{figure}
        \includegraphics[width=9cm]{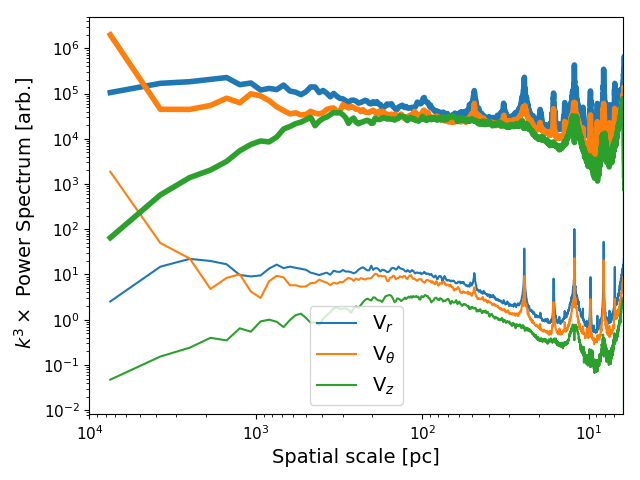}
        \caption{Power spectra of the three velocity components from face-on mass-weighted velocity projection maps. The spectra are compensated by $k^3$. The F65 model is shown with bold lines and the F10 model with thin lines. The F10 power spectra are shifted down by 3 dex for the sake of clarity. The spikes seen at the small spatial scales are due to oversampling because of the AMR grid out of which the Fourier transform was performed. These spectra show that the injection scale of turbulence is indeed different in the two models, and corresponds roughly to the disk scale height.} 
        \label{fig:PS_vel_nozoom}
    \end{figure}
    
   In Fig.~\ref{fig:PS_vel_nozoom} we show the three velocity component power spectra for both the F10 and F65 simulations, from the snapshots shown in Fig.~\ref{fig:star_surf_dens}. These power spectra are measured from 2D maps at the highest resolution. Thus, low resolution regions are interpolated. In the Appendix, we present power spectra with increasingly coarse resolution limit, down to the uniform grid limit. It is shown that this interpolation to higher resolution does not change the results. One can see several features in Fig.~\ref{fig:PS_vel_nozoom}:
   \begin{itemize}
       \item on large scales, there is a de-correlation between the vertical and planar velocities;
       \item on small scales, the three velocity components follow the same power-law, which differs between the F10 and F65 models.
   \end{itemize}
   
   The de-correlation between the vertical and the planar velocities was already obtained in previous works \citep[see e.g.][]{Bournaud2010} and observations \citep[see e.g.][]{Grisdale2017}. It was interpreted as being due to the injection scale for turbulence, thought to be similar to the disk height. We do reproduce this result here: the transition happens approximately at their disk scale height: 100~pc and 1~kpc for the F10 and F65 models, respectively.
   
   We fit power-laws to the velocity power spectra. As is done in the literature, we define the exponent of the power-law with respect to the norm of the wave vector $k$, which is the inverse of the spatial scale $l$: $k=1/l$. The two power-law exponents for small scales differ between the two models. For the radial, tangential and altitudinal velocity components, we obtain exponents -3.4, -3.3 and -3.0 for F10 and -4.4, -4.5 and -4.2 for F65. We note that the inertial range, which is defined as the range in spatial scales for which the velocity power spectrum is well fit by a single power law, is quite short for the F10 simulation, around 1 dex. A power-law with index equal to -3 corresponds to the scaling expected for a 2D power spectrum in \citet{Burgers1948} turbulence, compared to an index equal to -2.6 for \citet{Kolmogorov1941} turbulence.

\section{Results: turbulence cascade down to sub-pc}
\label{sec:results}

    In this Section, we present the effect of large scale gravitational instabilities on the structure of the ISM in the encapsulated zoom region. In what follows, if not stated otherwise, the analysis is performed on a single snapshot of the simulation 10~Myrs after the activation of the final level of refinement (level 20).

    \subsection{Gas distribution}
    \label{subs:dist}

    \begin{figure*}[h!]
        \centering
        \includegraphics[width=9.1 cm]{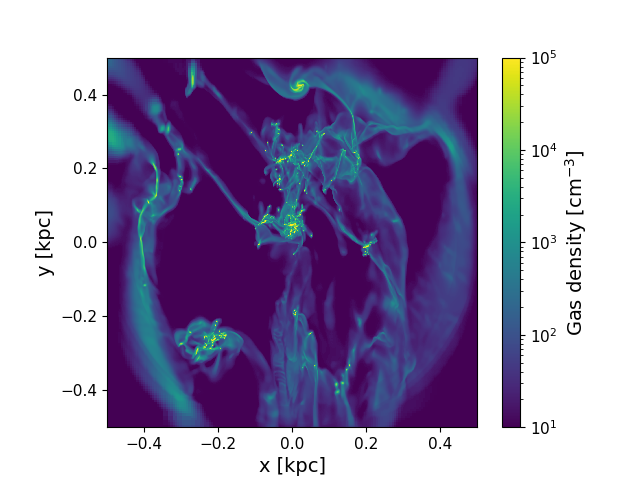}
        \includegraphics[width=9.1 cm]{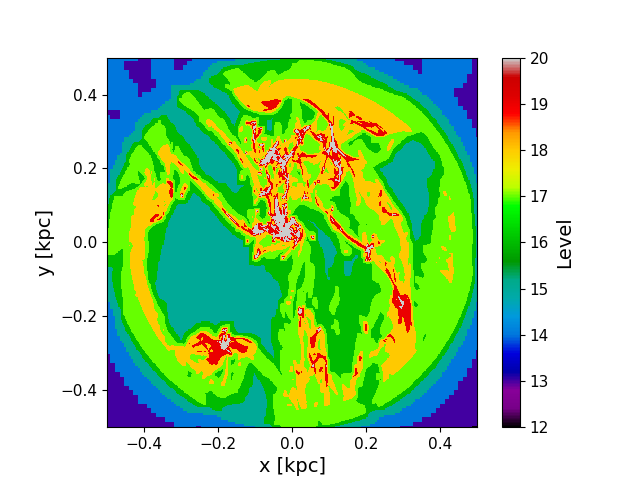}\\
        \includegraphics[width=9.1 cm]{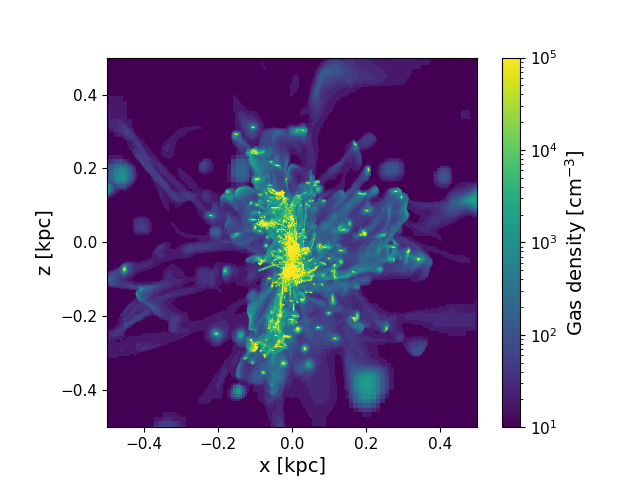}
        \includegraphics[width=9.1 cm]{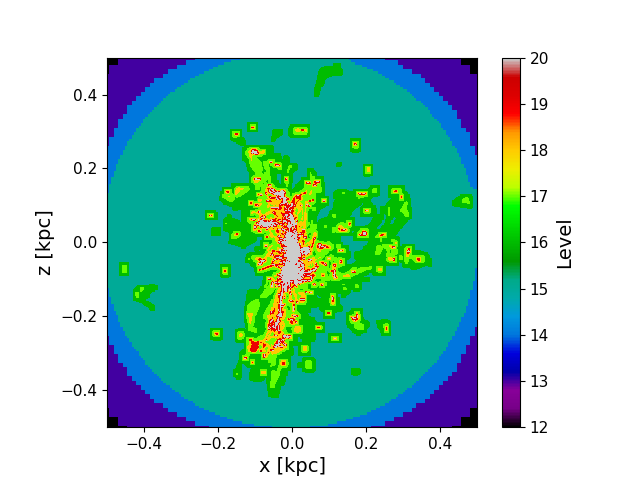}\\

        \caption{Left panels: gas density maps. Right panels: maximum level of refinement along the line-of-sight. Top panels: Zoom in the F10 simulation, seen face-on. Lower panels: zoom in the F65 simulation seen face-on. }
        \label{fig:zoom_rho}
    \end{figure*}

     \begin{figure}
        \centering
        \includegraphics[width=9cm]{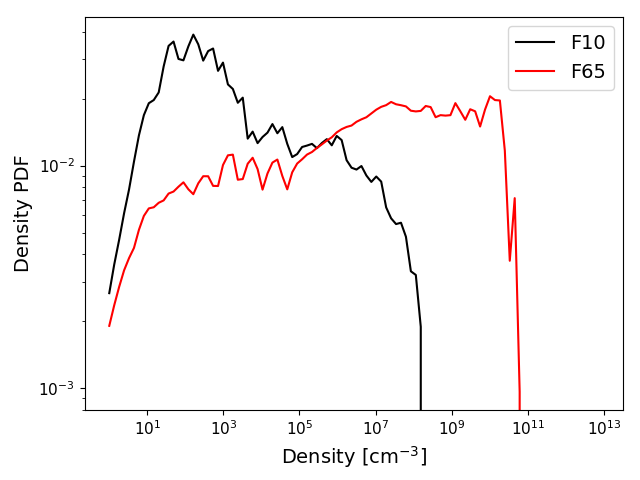}
        \caption{Normalized gas density PDF in the zoom regions in the F10 (black) and F65 (red) models 10~Myr of evolution after the activation of level 20.}
        \label{fig:zoom_PDF}
    \end{figure}

         \begin{figure}
        \centering
        \includegraphics[width=9cm]{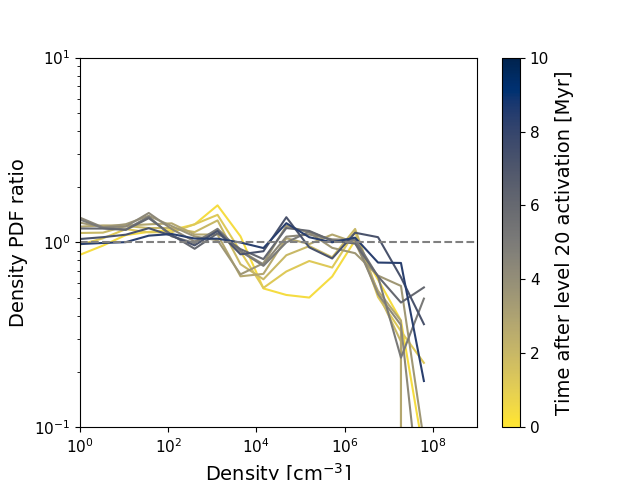}
        \includegraphics[width=9cm]{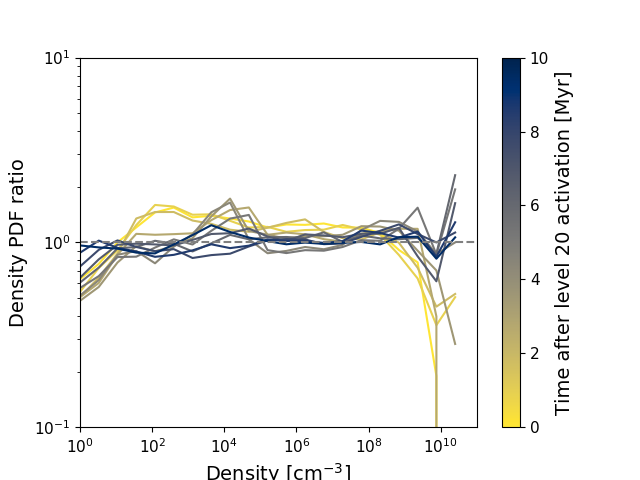}
        \caption{Evolution of the ratio between the gas density PDF to the final gas density PDF. The top (resp. bottom) panel is for the F10 (F65) simulation. We see that the density PDF converge to a stable distribution after a few Myrs.}
        \label{fig:zoom_PDF_evol}
    \end{figure}

        Gas density maps in the zoom region after 10~Myrs are shown in Fig.~\ref{fig:zoom_rho}. The gas fragments into very dense, n > $10^7 \,\rm H\,cm^{-3}$, gas clumps. The gas density PDFs in the zoom regions of each galaxy are shown in Fig.~\ref{fig:zoom_PDF}. In the F65 simulation, gas reaches densities of a few $10^{10} \,\rm H\,cm^{-3}$, almost three orders of magnitude above the highest densities reached in F10. In order to study the evolution of the gas density distribution after the activation of level 20, we show the ratio of the gas density PDF at different times with respect to the last snapshot, 10~Myr after the activation of level 20 in Fig.~\ref{fig:zoom_PDF_evol}. We see that for both models, the high-density end of gas density PDF increases progressively with time, and that after $\simeq 7$~Myr the gas density PDFs do not differ by more than a factor 2 per density bin.

        One should note that this convergence of the gas density PDF only happens thanks to the pressure support from the Jeans polytrope. Indeed, without this pressure floor at high density, gas would continue its collapse under the effect of its own gravity. Inclusion of this Jeans polytrope thus allows us to freeze the collapse of gas onto pressure supported sub-structures, analogous to {\it hot cores} in observed molecular clouds.

    \subsection{Gas turbulence}
    \label{subs:turb}
    
    \begin{figure}
        \centering
        \includegraphics[width=9cm]{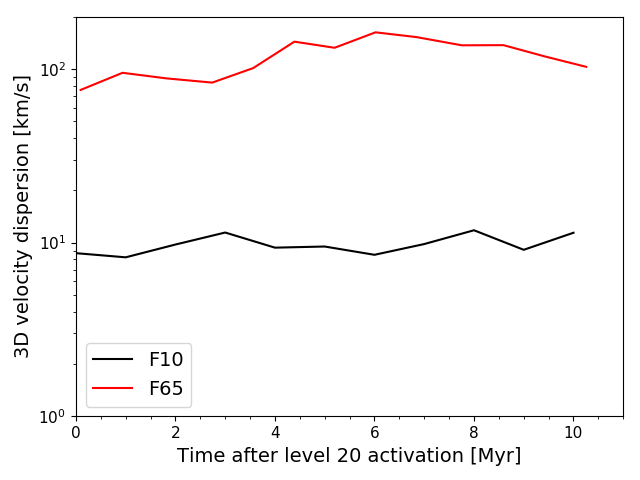}
        \caption{Evolution of the 3D velocity dispersion in F10 and F65 after the activation of level 20. There is not much evolution in 3D velocity dispersion over the 10~Myr after the activation of level 20. }
        \label{fig:mach_zoom}
    \end{figure}
    
    \begin{figure}
        \centering
        \includegraphics[width = 9cm]{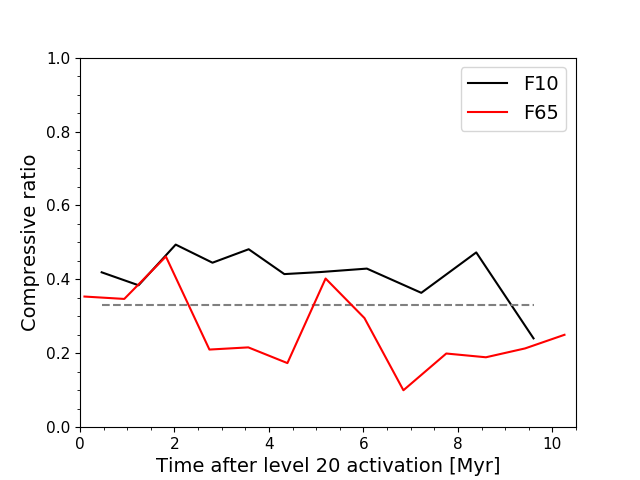}
        \caption{Evolution of the compressive ratio $\zeta$ after the activation of level 20 in F10 (black) and F65 (red) The dashed grey line shows the locus of energy equipartition: $\zeta = 30\%$. The evolution of the ratio is very noisy for the two runs, but remains consistent with a value close to equipartition.}
        \label{fig:zoom_comp_ratio}
    \end{figure}

    We measure the 3D velocity dispersion $\sigma_\mathrm{3D}$ in the zoom region at the uniform resolution of 3~pc (level 15) by mass-weighting the variances obtained in each parent cell at level 14, containing $2^3$ cells at level 15. We obtain 3D velocity dispersions of 11.4~km~s$^{-1}$ and 103.0~km~s$^{-1}$ for the F10 and F65 models, respectively. These velocity dispersion are of the same order as those observed in galaxies with corresponding gas fraction \citep[see review by][]{Forster2020} and correspond to supersonic gas, relative to the ambient isothermal sound speed: 1.8~km~s$^{-1}$ and 11.7~km~s$^{-1}$ for a mean molecular weight of 0.6 and a temperature of 236~K and $10^4$~K, respectively.
    
    The evolution of \sthreeD~after the last level activation is shown in Fig.~\ref{fig:mach_zoom}. It evolves towards a stabilized value after $\simeq$~6~Myr in each simulation. This convergence time is similar to the convergence time of the gas density PDF presented in the previous sub-section.
    
    We then study how these turbulent motions are structured. In Fig.~\ref{fig:zoom_comp_ratio} we show the compressive ratio $\zeta$, defined as \citep[see e.g.][]{Kida1990,Kritsuk2007}: 
    \begin{equation}
        \zeta = \frac{ <|\nabla \cdot \vec{v}|^2>}{ <|\nabla \cdot \vec{v}|^2> + <|\nabla \times \vec{v}|^2>}\, ,
    \end{equation}
    where the average are is weighted by mass, to get energy ratios. Following the convergence criterion of \citet{Renaud2015}, we compute this ratio for cells and parent cells that are larger than 8 times the highest resolution cells, that is above or equal to level 17 \citep[see][]{Grisdale2017}. This is done to correctly capture solenoidal motions which need more resolution elements to be resolved than compressive motions \citep{Federrath2011}.

    The evolution of this ratio is rather noisy, typically between 20 and 40\% with rapid variation between snapshots. At energy equipartition, one expects $\zeta = 33\%$ from a dimensional argument (compressive motions are along one direction, whereas solenoidal motions are along two dimensions, see e.g. \citealt{Hennebelle2012}). During the first 10~Myrs after the activation of level 20, $\zeta$ is typically above that value in F10, and below it in F65. However, given the noise in the measurements, the data does not allow us to conclude that compressive motions are stronger in either of the simulations. 
    
    \subsection{Velocity structure}
    \label{subs:ps_v}
    
    \begin{figure}
        \includegraphics[width=9cm]{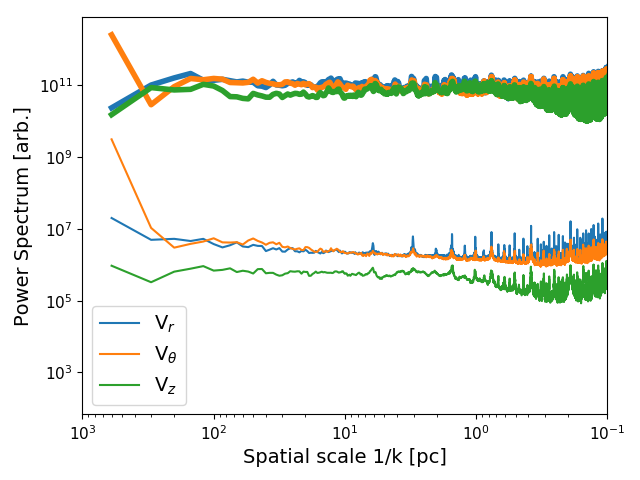}
        \includegraphics[width=9cm]{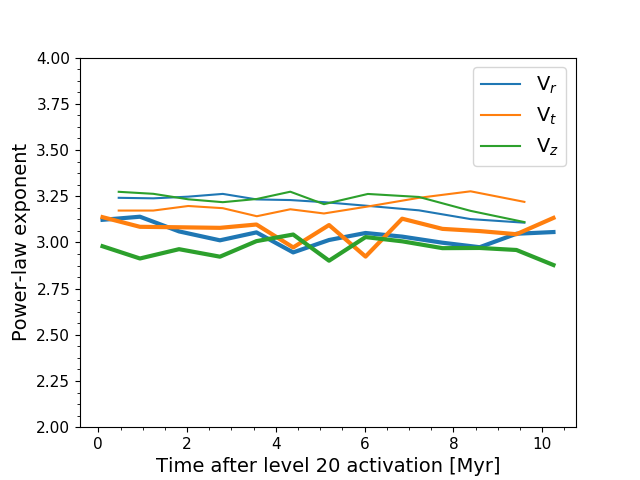}
        \caption{Top panel: 2D power spectrum of the three velocity components 10 Myr after the activation of level 20. The curves are compensated by $k^3$. Curves from to the F65 run are in bold, while thin lines are for the F10 run. The F10 power spectra are shifted down by 3 dex for the sake of clarity. Bottom panel: evolution of the power-law index of the fit of the velocity power spectra. Curves related to the F65 run are in bold. The power spectra are well fitted by a power-law with index close to -3, and there is not much evolution during the 10~Myr after the activation of the level 20.}
        \label{fig:zoom_PS_vel}
    \end{figure}

    Energy transfer from large to small scales via turbulence is imprinted into the velocity power spectrum. In Fig.~\ref{fig:zoom_PS_vel} we show the velocity power spectra for the three velocity components and their evolution after the activation of level 20. We see that the power-spectra are well fitted by power-laws between the spatial scales of 100~pc and 1~pc. Below this range, we see a steepening of the power spectra. This will be discussed in Section~\ref{sec:disc:burgers}. Moreover, we see a flattening of the power spectrum of $v_z$ for the F10 run at a spatial scale of 100~pc. This is similar to what was observed in Fig.~\ref{fig:PS_vel_nozoom} and is interpreted as the transition between 2D and 3D turbulence at the disc scale height. 
    
    We fit power laws to the 2D velocity power spectrum. The fits are performed between 1~pc and 100~pc for F10 and 1~kpc for F65. or the radial, tangential and altitudinal velocity components, we obtain exponents -3.1, -3.2 and -3.1 for F10 and -3.0, -3.1 and -2.9 for F65.
    They are similar for each galaxy model and close to the value of -3, which is the scaling of Burgers' turbulence. In the F65 simulation, the injection scale is around 1~kpc, that is the disk scale height, seen as a transition in Fig.~\ref{fig:PS_vel_nozoom}, and we follow the same power-law down to scales of around 1~pc. Thus our setup is able to resolve a turbulent inertial range over three orders of magnitude in spatial scales, which is more than the largest supersonic turbulence simulations so far \citep[see e.g.][with an inertial range of 2 orders of magnitude]{Federrath2020}.

    In Fig.~\ref{fig:zoom_PS_vel},  one can note that the $v_z$ power spectrum for the F10 simulation lies between 0.5 and 1 dex below the $v_r$ and $v_\theta$ power spectra. The $v_z$ power spectrum for the F65 lies also below the $v_r$ and $v_\theta$ power spectra, but only by at most 0.3 dex. This effect can also be seen in Fig~\ref{fig:PS_vel_nozoom}, in a lesser extent. This is interpreted as a consequence of the stronger anisotropy of the velocity field between in-plane and vertical motions in the F10 galaxy: in the zoom region maps used to compute the power spectra, the standard deviation of the $v_z$ distribution is around three times lower than that of $v_r$ and $v_\theta$ in the F10 galaxy, but only up to 50\% lower in the F65 galaxy. These different anisotropies could have originated from the fact that the gas in the F10 simulation is dominated by the gravity of the stellar background, while in the F65 galaxy, the self-gravity of the gas is dominant. Indeed, in the zoom region the gas mass fraction is 14~\% and 87~\% in the F10 and F65 galaxies, respectively. A detailed study of the anisotropy of the velocity field, and its origin, is out of the scope of the present paper.
    
    \subsection{Gas structure}
    \label{subs:structure}
    
    \begin{figure}
        \includegraphics[width=9cm]{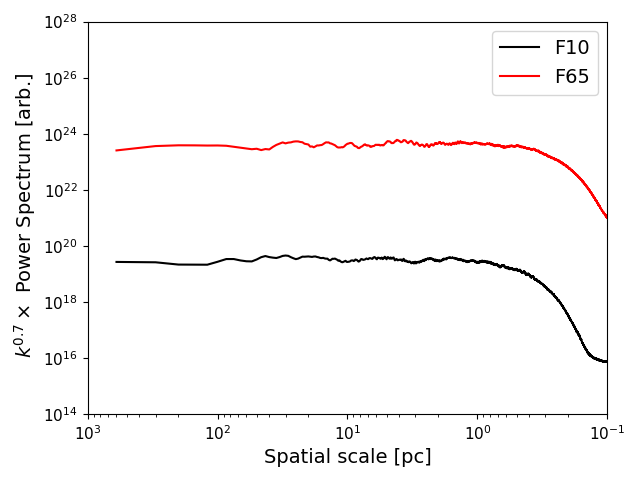}
        \includegraphics[width=9cm]{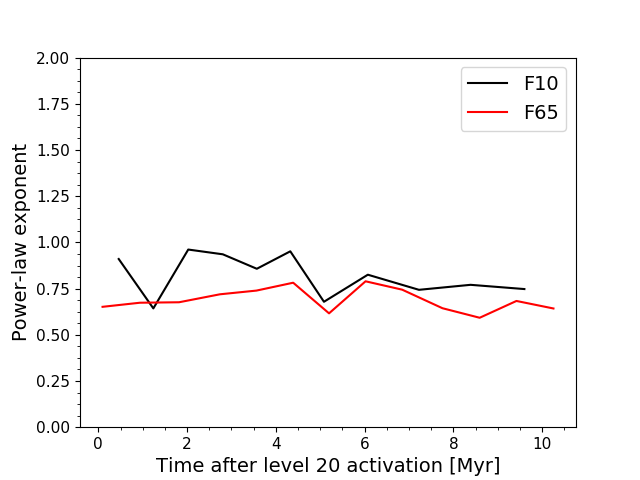}
        \caption{Top panel: Surface density power spectra for F10 and F65, 10~Myrs after the activation of level 20. The curves are compensated by $k^{0.7}$. Bottom panel: evolution of the power-law of the fit of the velocity power spectra. The power spectra are well fitted by a power-law with index close to -0.7, and there is not much evolution during the 10~Myr of the simulation.}
        \label{fig:zoom_ps_rho}
    \end{figure}
    
     \begin{figure}
        \centering
        \includegraphics[width=9cm]{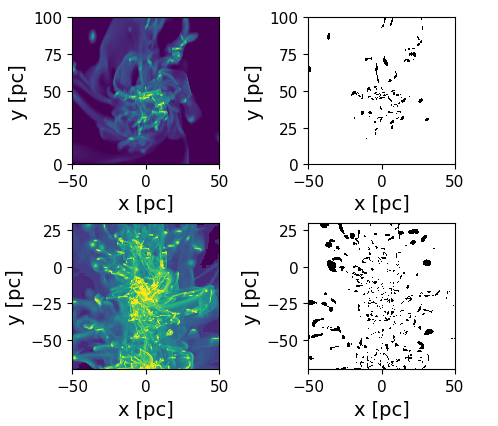}    
        \caption{Left panel: Central regions in the zoom. The origin is the center of the zoom region. Right panel: gravitationally bounds clumps (with $\alpha_\mathrm{vir} < 2$, see text). The top and bottom panels show the F10 and F65 runs, respectively.}
        \label{fig:astrodendro}
    \end{figure}

    \begin{figure}
        \centering
        \includegraphics[width=9cm]{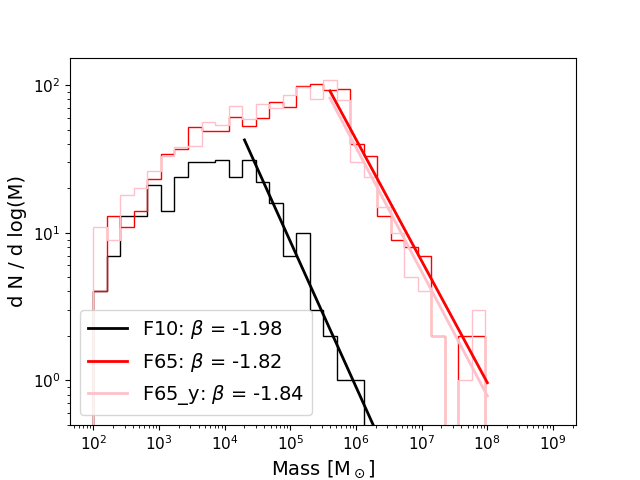}
        \includegraphics[width=9cm]{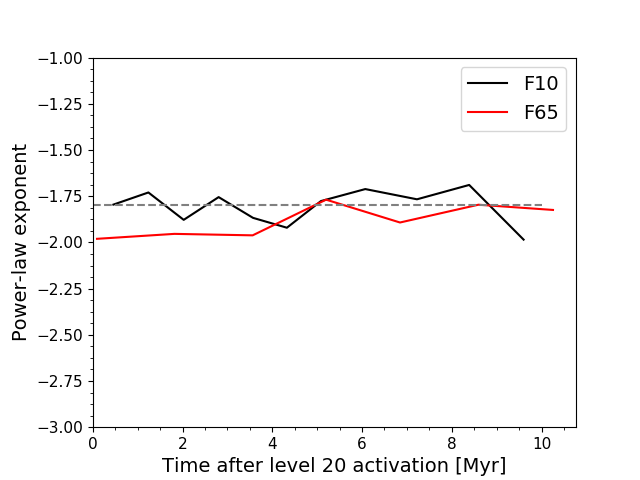}
        \caption{Top panel: Mass spectrum of the bound structures in the zoom ($\alpha_\mathrm{vir} < 2$). The pink histogram and fit are obtained for a detection performed on the edge-on density map. The power-slopes are parametrised by $dN/dM \propto M^{-\beta}$. Bottom panel: Evolution of the slope of the power-law fit to the high-mass end of the mass spectrum. The dashed grey line shows the location of the -1.8 exponent. The slopes are very similar for the two models and do not evolve much during the 10~Myrs of the simulation.}
        \label{fig:mass_spec}
    \end{figure}

Turbulence creates structures in the gas. One may wonder if these structures differ with the gas fraction. For instance, both analytical \citep{Saichev1996} and numerical simulations of compressible non-gravitating gas \citep{Kim2005, Kritsuk2007, Konstandin2016} find density power spectra following power-laws with indices going from -2 for sub-sonic turbulence, to 0 at the infinite Mach number limit. In Fig.~\ref{fig:zoom_ps_rho} we show the power spectra of the gas surface density in the zoom region for each simulation. We see that the power spectra are strikingly similar, despite big differences in the density maps seen in Fig.~\ref{fig:zoom_rho}. In particular, at spatial scales higher than one parsec, the power spectra are well fitted by a power-law with exponent of $\simeq -0.7$. We note that this power law is a good fit during the full 10~Myr of evolution after the activation of level 20. Furthermore, both power spectra get steeper at scales smaller than 1~pc, similar to the velocity power spectra shown in Fig.~\ref{fig:zoom_PS_vel}. This result is rather surprising, given that several studies have shown a dependence of the power-law exponent on the Mach number, combined with the fact that the Mach number is different in the two simulations (from 6.5 to 9, see Section~\ref{subs:turb}). Note that those studies did not include self-gravity.

To study further the density structure inside the zoom region, we use the python package {\sc astrodendro}\footnote{accessible at this address: https://github.com/dendrograms/astrodendro/} to detect bound sub-structures. We detect all structures with a peak density above $10^4\, \rm H\, cm^{-3}$ and with an area of at least 20 pixels square on the density map, that is 0.18~pc$^2$. We note that this density threshold is lower than the one at which cells are affected by the pressure support from the Jeans polytrope (see Section~\ref{subs:method}). We then compute the virial parameter $\alpha_\mathrm{vir}$ of each clump, defined as \citep[see e.g.][]{Bertoldi1992}: 

\begin{equation}
    \alpha_\mathrm{vir} = \frac{5}{3} \frac{\sigma_\mathrm{3D}^2~R_{1/2}}{\mathrm{G} M}\, ,
\end{equation}

where $R_{1/2}$ is the half-mass radius of the clump, $M$ its gas mass and $\sigma_\mathrm{3D}$ its internal 3D velocity dispersion. In order to account for the non-homogeneity and non-sphericity of the clumps we use the half-mass radius instead of the detection radius, and we consider that a clump is gravitationally bound if $\alpha_\mathrm{vir} < 2$. The following results are not qualitatively affected by using  $\alpha_\mathrm{vir} < 1$ or $\alpha_\mathrm{vir} < 5$, as will be described in the following. Examples of gravitationally bound clumps in both simulations are shown in Fig.~\ref{fig:astrodendro}.

There are several hundreds of such clumps in each simulation. The typical clump size is around 1~pc, which corresponds to the spatial scale of the breaks of power spectra seen in Fig.~\ref{fig:zoom_PS_vel} and \ref{fig:zoom_ps_rho}, both of them being steeper for smaller scales. We stress that the physics of the gas for our clumps is not realistic, in the sense that we stop the isothermal collapse by introducing a pressure floor, and we do not remove the gas into star formation, and thus have a continuous accretion of matter onto our sub-structures, which affects the clump sizes. The mass spectra of the clumps are shown in Fig.~\ref{fig:mass_spec}. We see that, for the F65 simulation, using edge-on or face-on maps does not modify significantly the number and masses of the gravitationally bound clumps. The last two dex of the high-mass end of the mass spectra are well fitted by power-laws with an index close to -1.8. These power-laws vary by less than 3 \% around these values if one changes the selection criterion to $\alpha_\mathrm{vir} < 1$ or $\alpha_\mathrm{vir} < 5$. One should note that the hierarchy of structures in a turbulent field without gravity produces a mass spectrum with index -2 \citep{Elmegreen2009} whereas numerical studies by \citet{Hennebelle2007, Audit2010}, which did include different gas cooling functions or equations of state, but not self-gravity, found a similar slope of -1.7.  It should be noted that this exponent -1.8 is the same as that of CO clumps mass distributions \citep[in][see section 4.3]{Hennebelle2012}. Thus this scaling seems then to be universal for any turbulent medium, regardless of the nature of the drivers.

\section{Discussion}
\label{sec:discussion}

    \subsection{Origin of Burgers' turbulence scalings}
    \label{sec:disc:burgers}

In Section~\ref{subs:ps_v} we obtained 2D velocity power spectra well fitted by a power law with index around -3, which is similar to the Burgers' turbulence scaling. Burgers' turbulence is a model where the transfer of energy to small scales is achieved via shocks. This is different from the Kolmogorov turbulence model, which applies to incompressible fluids and for which the transfer of energy to small scales is achieved via eddies. In this model, the 2D velocity power spectrum follows a power law with index -8/3 \citep{Kolmogorov1941}.

The inviscid 3D Burgers equation is pure advection and reads:
\begin{equation}
    \frac{\partial}{\partial t} \vec{v} + (\vec{v} \cdot \nabla) \vec{v} = 0\, .
    \label{burger}
\end{equation}

Our simulation uses an isothermal equation of state up to $10^6$ and $10^{7.5}\,\rm H\,cm^{-3}$ at the resolution limit, for F10 and F65, respectively. The regions affected by this pressure floor are mainly located in the gravitationally bound sub-structures (see above). Thus our effective Euler equation, outside pressure floor affected regions, and with $\phi$ is the gravitational potential, reads:
\begin{equation}
    \frac{\partial}{\partial t} \vec{v} + (\vec{v} \cdot \nabla) \vec{v} = -\frac{1}{\rho} \nabla P -\frac{1}{\rho} \vec{\nabla}\phi\, .
\end{equation}

In the supersonic case, where pressure does not affect much the motions, one expects the velocity power spectrum to follow the scalings of Burgers turbulence for 1D and 2D turbulence \citep[see e.g.][]{Boldyrev2002}. In 3D, the scaling might change because of vorticity generation \citep[see also][]{Hennebelle2007, Padoan2016, Iffrig2017}. 

Without viscosity  and with gravity as the only external force, the evolution of the vorticity $\vec{\omega}  = \nabla \times \vec{v}$ reads:
\begin{equation}
    \frac{\partial \boldsymbol \omega}{\partial t} + (\mathbf v \cdot \nabla) \boldsymbol \omega = (\boldsymbol \omega \cdot \nabla) \mathbf v - \boldsymbol \omega (\nabla \cdot \mathbf v) + \frac{1}{\rho^2}\nabla \rho \times \nabla P + \nabla \times \left( \frac{ - \nabla \phi}{\rho} \right)\, . 
    \label{eq:vorticity}
\end{equation}

In our isothermal setup, $\nabla P \propto \nabla \rho$, and as the gravitational force is a conservative force, it does not generate vorticity. This can also be seen in Fig~\ref{fig:zoom_comp_ratio}, where we show that the compressive ratio stay close to equipartition during the full 10~Myr of the simulation. Furthermore, the Jeans polytrope mainly affects gravitationally bound substructures which have a typical size of 1~pc. This explains why the 2D velocity power spectra are well fitted by power-laws with an exponent close to -3 for scales between the injection scale and the start of the pressure support, as seen in Fig.~\ref{fig:zoom_PS_vel}. We note that this scaling is not likely to be found in observations if the velocity field is impacted by forces generating vorticity.

    \subsection{Importance of other physics}
    \label{sec:disc:physics}

    Our numerical setup only accounts for hydro-dynamics and self-gravity. In particular we impose a constant temperature (except in the highest density cells). One may wonder what would change if more physics, such as cooling, magnetic fields or star formation and its feedback, were introduced.
    
    \subsubsection{Cooling}
    
    First, if a full cooling and heating model was implemented, we would have pressure gradients in the Euler equation. For non-barotropic fluids, the term $\nabla \rho \times \nabla P$ would not necessarily vanish and could induce vorticity. A deviation from Burgers' scaling would then be possible \citep[see e.g.][]{Hennebelle2007, Padoan2016}. 
    
    There are two stable phases at pressure equilibrium in the ISM: the {\it warm neutral medium} at T $\simeq 10^4$~K and the {\it cold neutral medium}, at T $\simeq 100$~K. The transition between these two stable phases is a thermal instability due to an elevation of the local density. Thus it happens at a characteristic spatial scale, depending on the average gas density and metal content of galaxies. Such a process could then leave its imprint in the turbulent cascade. This characteristic scale depends on the average gas density in galaxies, may vary with the gas fraction, unlike the gravity-driven turbulence presented above. 

    \subsubsection{Magnetic fields}

    Magnetic fields may affect the turbulence cascade in two ways. 
    
    First, magnetic fields act as a source term in the vorticity equation. In the presence of a magnetic field, equation~\ref{eq:vorticity} has an additional term on the right hand side corresponding to the rotational of the Lorentz force: $\nabla \times (\vec{J}\times\vec{B})/\rho$, with $\vec{J}$ and $\vec{B}$ the current and the magnetic field, respectively. This creation of vorticity may impact the transfer of energy by creating eddies, and thus change the velocity power spectrum \citep[see e.g.][]{Hennebelle2007, Padoan2016}.

    Second, magnetic fields may also induce turbulence via the magneto-rotational instability \citep[see e.g.][]{Kim2003, Piontek2007}. This instability occurs in disks in the presence of a poloidal component of the magnetic field in a region where the angular velocity decreases with galactocentric radius, which happens for instance in the flat velocity rotation part of galaxies.

    Thus, magnetic fields could in principle leave their imprint on the isothermal turbulence cascade in galactic disks, and their study would necessitate dedicated numerical simulations.
    
    \subsubsection{Star formation and feedback}
    
    First, enabling star formation would allow gas depletion in the densest cell. In the long term, star formation should reduce the gas fraction of the galaxy. However, we have seen that the characteristics of the turbulent cascade do not change with the decrease of gas fraction, thus we do not expect significant changes in our results if we formed stars instead of using the Jeans polytrope to stop the fragmentation at the limit of resolution.
    
    Second, star formation feedback, in the form of heating and radiative pressure from HII regions and supernovae (SN) explosions would also inject energy at a given scale, likely around 50~pc to 100~pc with a dependence on SN - and thus star formation - clustering but also ambient density gas \citep{Padoan2016, Iffrig2017, Ostriker2022}. The fraction and spatial scale of energy deposited in the ISM may also be a function of disk scale height and thus gas fraction \citep{Orr2022a, Orr2022b}. By releasing kinetic energy at a characteristic spatial scale, star formation feedback may thus leaves an imprint on the turbulent cascade. 
    
    Several works on turbulence in the ISM either focused on galaxy scales, without forcing \citep[see e.g.][]{Bournaud2010, Falceta2015, Grisdale2017, Kortgen2021, Bieri2022}, or on ISM boxes typically 1~kpc$^3$, thus without the self-consistent energy injection from large scales \citep[see e.g.][]{Iffrig2017, Brucy2020, Hu2022,Rathjen2021}. In either case, inclusion of a thermodynamical model and star formation feedback gives velocity power spectrum scalings between that of Burgers and \citet{Kolmogorov1941}, which was primarily obtained for incompressible turbulence.
    
    Thus, the inclusion of both cooling and star formation and its feedback may leave imprint of their characteristic scales onto the turbulent cascade, a scale which may depend on the gas fraction of the galaxy. In a forthcoming paper, we include these processes to quantify their effects on the turbulent cascade.

    \subsection{Comparison to observations}
    
    Observations of the gas and velocity structure in the atomic ISM are mostly achieved via the 21-cm HI line. Given its difficulty, Fourier analysis of the gas structure is done achieved only for the surface density and not the line-of-sight velocity \citep[see e.g.][]{Elmegreen2001,Dutta2008, Grisdale2017}. The velocity power spectrum for galaxies may be achieved using other tracers, such as ionised gas, or stars in the solar neighborhood \citep{Bovy2015}.
    
    \citet{Grisdale2017} computed the 2D power spectra of HI gas in six galaxies from the THINGS survey \citep{Walter2008} and compare them to their simulations, which include gas cooling and star formation. They observe HI power spectra that are much steeper than ours: they are fitted with power laws with slopes varying from -1.6 to -2.8, agreeing with their simulations, whereas we have found an universal isothermal slope of -0.7. Theory and simulations predict power laws with slopes between 0 and -2 for isothermal sub-sonic to supersonic compressible non-gravitating gas (see Section~\ref{subs:structure}).

    One should note that analysis of the observational data was limited by the large beam size, between 100 and 300~pc, which is the same order of magnitude as the disk scale height. At those scales, turbulence acts on two dimensions, instead of three as in our setup, which could change the index of the power spectrum. However, the resolution limit of the simulations was much lower (4.6~pc) and the their surface density power spectra are measured down to $\simeq 50$~pc. We note that their fits are all steeper than ours, even for their shallower power spectra, obtained in the case where they do not have stellar feedback. One may thus wonder if gas cooling and star formation feedback would not be the cause of the steepening of HI power spectra. This will be studied in a forthcoming paper.

\section{Conclusions}
\label{sec:conclusions}

The aim of this paper is to study the characteristics of isothermal gravity-driven turbulence in galaxies with two different gas fractions. We follow the turbulence cascade from its injection scale (100~pc to 1~kpc depending on the galaxy model) down to the resolution limit of 0.095~pc, using a zoom-in method on a gas overdensity. This method allows us to probe self-consistently the self-generated turbulence cascade over three orders of magnitude in spatial scales, and over 10~Myrs.

As expected, the difference in gas fraction triggers different types of instabilities, and thus galaxy morphologies. The F10 model develops spiral arms while the F65 disk fragments into many massive gas clumps. Despite very different morphologies in the gas distribution and structure, and the order of magnitude difference in the spatial scale of turbulence generation, we find that the turbulence and gas structure cascade follows the same scalings laws in both setups.

In particular:
\begin{itemize}

    \item the velocity power spectrum follows the Burgers' scaling,
    \item the surface gas density power spectrum has a power-law slope of -0.7,
    \item gravitationally-bound substructures follows a mass distribution with -1.8 slope, similar to that of CO clumps. 
\end{itemize}

We note that the turbulent velocity and density fields reach a steady-state after $\simeq 6\,\rm Myr$. This time scale is short compared to the lifetime of giant molecular clouds, of a few tens of Myrs \citep[see review by][]{Chevance2022}, in particular shorter than the time it takes internal feedback to have a significant effect. Thus, this setup provides insights on plausible initial conditions for isothermal and non-magnetic star-forming clouds \citep[see also][]{Lane2022}. 

These simulations thus suggest a universality of gravity-driven isothermal turbulent cascade in galaxies across cosmic times. Our encapsulated zoom method is a promising tool to study the interplay between turbulence injection processes at widely different spatial scales.

\begin{acknowledgements}
We thank the anonymous referee for their careful reading and detailed comments which improved the paper. We thank Eva Ntormousi and Elliot Lynch for very stimulating discussions. Simulations were produced using PRACE (grant 2020225366) and GENCI allocations (grants A0090411111 and A0110411111). We gratefully acknowledge support from the PSMN (Pôle Scientifique de Modélisation Numérique) of the ENS de Lyon for the computing resources. JF, FB, NB and PH acknowledge support from PNCG. NB and PH acknowledge financial support from the European Research Council (ERC) via the ERC Synergy Grant "ECOGAL: Understanding our Galactic ecosystem -- From the disk of the Milky Way to the formation sites of stars and planets" (grant 855130). We made use of Astropy (Astropy Collaboration et al. 2013), with heavy usage of the Python packages NumPy (Walt et al. 2011), iPython (Prez \& Granger 2007), SciPy (Jones et al. 2001), matplotlib (Hunter 2007). We would thus like to thank all who designed and contributed to these excellent pieces of software, and most importantly made them available to the community.
This research made use of astrodendro, a Python package to compute dendrograms of Astronomical data (http://www.dendrograms.org/).
\end{acknowledgements}

\bibliographystyle{aa}   
\bibliography{library}

\begin{appendix}

\section{Computing 2D power spectra on non-uniform grids}

\begin{figure}
        \centering
        \includegraphics[width=9cm]{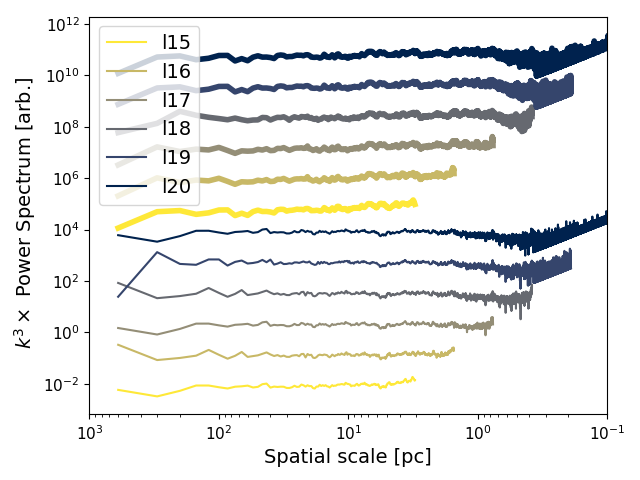}
        \includegraphics[width=9cm]{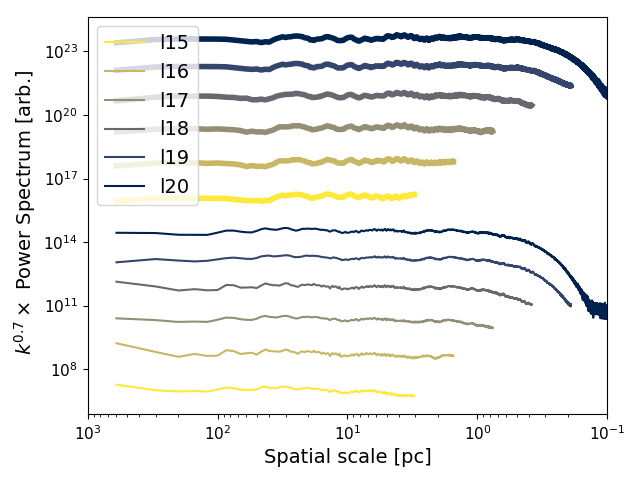}
        \caption{Compensated power spectra computed with on 2D maps for the maps shown in Fig.~\ref{fig:zoom_rho}. The upper and lower panels show altitudinal velocity and surface density power spectra, respectively. Each color corresponds to a different highest resolution of the cells, varying from level 20 down to level 15, which corresponds to a uniform grid. Power spectra in bold and thin lines correspond to the F65 and F10 runs, respectively. The latter are all shifted down by 5~dex for the sake of clarity We do not see a qualitative difference between the different levels. }
        \label{fig:ps_conv}
    \end{figure}

Throughout the paper, power spectra are measured on 2D maps at the highest resolution. Thus, low resolution regions are interpolated. In this section, we present power spectra measured with coarser and coarser highest resolution, down to the uniform grid limit, at the level 15. The results of the tests, both for altitudinal velocity and surface density power spectra, are shown on Fig.~\ref{fig:ps_conv}. We see that the slope of the compensated power spectra is not changed by the fact that we use cells with different resolution.

\end{appendix}

\end{document}